\documentclass[reprint, amsmath,amssymb, aps]{revtex4-2}
\usepackage[utf8]{inputenc}
\usepackage{caption}
\usepackage{subcaption}
\usepackage{fullpage}
\usepackage{xcolor}
\usepackage[titletoc]{appendix}
\usepackage{floatrow}

\usepackage{verbatim}
\usepackage{lineno}
\usepackage{placeins}
\usepackage{natbib}
\usepackage{graphicx}
\usepackage{indentfirst}
\usepackage{hyperref}
\usepackage{upgreek}
\usepackage{multirow}
\usepackage{makecell}
\usepackage{booktabs}
\usepackage{xspace}
\usepackage{titlesec}
\titleformat{\section}{\normalfont\normalsize\bfseries\center} {\thesection}{1em}{}
\titlespacing\section{0pt}{12pt plus 4pt minus 2pt}{10pt plus 2pt minus 2pt}
\titlespacing\subsection{0pt}{10pt plus 4pt minus 2pt}{8pt plus 2pt minus 2pt}
\titlespacing\subsubsection{0pt}{8pt plus 4pt minus 2pt}{8pt plus 2pt minus 2pt}
\usepackage{enumitem}
\setlist{nosep} 

\usepackage{siunitx}[=v2]
\usepackage[T1]{fontenc}

\usepackage[scaled=.91]{helvet} 

\newcommand{\SID}{{SiD}\xspace}
\newcommand{\KPIX}{{KPiX}\xspace}
\newcommand{\pT}{\ensuremath{p_T}}

\newcommand{\ee}{\ensuremath{e^+e^-}}

\newcommand{\julia}{\textsc{Julia}\xspace}
\newcommand{\python}{\textsc{Python}\xspace}

\newcommand{\cplusplus}{\textsc{C++}\xspace}
\newcommand{\fortran}{\textsc{Fortran}\xspace} 
\newcommand{\go}{\textsc{Go}\xspace} 
\newcommand{\rust}{\textsc{Rust}\xspace} 
\newcommand{\geant}{\textsc{Geant4}\xspace}
\newcommand{\lcio}{\textsc{LCIO}\xspace}

\newcommand{\CCC}{\ensuremath{\mathrm{C}^3}}

\begin{document}

\title{The \SID Detector concept \\Input to the European Strategy Process Update 2026}

\author{M.~Breidenbach}
\affiliation{SLAC National Accelerator Laboratory, 2575 Sand Hill Road, Menlo Park, CA, USA}
\author{J.E.~Brau}
\affiliation{Department of Physics, University of Oregon, Eugene, OR 97403, USA}
\author{P.~Burrows}
\affiliation{Department of Physics, Oxford University, Oxford, UK }
\author{T.~Markiewicz}
\affiliation{SLAC National Accelerator Laboratory, 2575 Sand Hill Road, Menlo Park, CA, USA}
\author{C.~Vernieri}
\affiliation{SLAC National Accelerator Laboratory, 2575 Sand Hill Road, Menlo Park, CA, USA}
\author{M.~Stanitzki}
\affiliation{DESY, Notkestrasse 85, 22607 Hamburg, Germany}
\author{J.~Strube}
\affiliation{University of Oregon, Institute for Fundamental Science, Eugene, OR 97403-5203}
\author{A.P.~White}
\affiliation{University of Texas Arlington, Arlington, TX 76019, USA}
\author{H.~Yamamoto}
\affiliation{Tohoku University, Sendai, Japan}

\begin{abstract}

The \SID Detector is one of two detector designs validated in 2012 for the  International Linear Collider (ILC).
\SID features a compact, cost-constrained design for precision Higgs and other measurements, with
sensitivity to a wide range of possible new phenomena. A robust silicon vertex and tracking system,
combined with a five Tesla central solenoidal field, provides excellent momentum resolution. The
highly granular calorimeter system is optimized for Particle Flow application to achieve very good
jet energy resolution over a wide range of energies. Given the advances in detector technology and the current set of three linear collider concepts under consideration,
the \SID team is reviewing its earlier design and technology decisions and
updating the design and choices with recent technological advances. For each area of \SID development R\&D topics and opportunities for participation will be discussed.
\clearpage
\end{abstract}

\maketitle

\newpage

\section{Introduction}

\setcounter{page}{1}

The International Linear Collider (ILC)~\cite{Behnke:2013xla} is a proposed \ee energy frontier collider,
based on superconducting cavities in a \SI{20}{\kilo\meter} long linear accelerator,
with an initial baseline center-of-mass energy of 250~GeV. The ILC will provide polarized beams for both electrons (80\%) and positrons (30\%), 
which is a unique capability of linear colliders. The ILC project includes a clear upgrade path to center-of-mass energies of 1~TeV, or even slightly beyond. 
The ILC has a mature baseline design  summarized in the 2012 Technical Design Report (TDR)~\cite{Adolphsen:2013jya,Adolphsen:2013kya}. 
While there have been technical advances, unfortunately, there has been little progress towards an international agreement to build 
the ILC. Two other designs have emerged that promise higher gradients and lower costs. 

The Compact Linear Collider (CLIC)~\cite{Linssen:2012hp,CLICdp:2018cto} has been proposed largely by CERN. It features normal conducting copper linacs with their RF power 
produced by decelerating a drive beam, and would have gradients up to ~100 MeV/m. CLIC is intended to be built in three stages, 
from 380 GeV to 3 TeV center-of-mass energies.

The Cool Copper Collider (\CCC)~\cite{vernieri2023cool,nanni2023status} has been proposed largely by SLAC. It features normal conducting copper linacs at operated at $\approx$ 80~K, 
with an advanced cavity design and using distributed coupling to feed RF to each cavity. 
It would operate at gradients of 70 and 120 MeV/m. Its first stage is 8~km long, starting  at 250~GeV and extending to 550~GeV center-of-mass energy with the addition of RF sources. It would extend to 3~TeV with additional length and higher gradients up to 155 MV/m.

The linear collider environment is unique and very different from synchrotrons. All of the linear colliders provide polarized electrons with the option of polarized positrons.  They all have small duty cycles with trains of bunches repeating at 5~Hz (ILC), 50~Hz (CLIC) and 60~Hz (\CCC). The trains consist of several hundred bunches with spacing ranging from ~300 ns (ILC) to a few ns for CLIC and \CCC. This allows buffering of the data in the front-ends during the train, and read out at the end of the bunch train.  Front end power would be limited between trains, reducing the average power consumption by roughly a factor of 100 to 1000. ILC would have slightly different data buffering schemes from CLIC and \CCC, but the differences seem quite tractable.

\SID started as a detector concept for linear colliders almost twenty years ago~\cite{Abe:2001rdr,Brau:2004ai}. It was well documented in the ILC TDR 
Detailed Baseline Document (DBD)~\cite{Behnke:2013lya} in 2012. 
This note will first give a brief review on the current design and layout of \SID and then identify and highlight the improvements appropriate 
for a construction start in the late 2020s, and the new opportunities for R\&D contributions. This note will not recapitulate the DBD in great details, 
and the reader should refer to the ILC~TDR for a complete summary of the physics motivations~\cite{Baer:2013cma}, the ILC accelerator~\cite{Adolphsen:2013kya} 
and the conceptual detector designs\cite{Behnke:2013lya}. For a review of the R\&D activities in the Linear Collider Community, the Detector R\&D Report~\cite{janstrube_maximtitov_2021} 
is an excellent summary. 

\section{\SID Status for the DBD}

The \SID detector has been designed as a general-purpose experiment designed to perform precision measurements
at the ILC. It satisfies the challenging detector requirements resulting from the full range of 
ILC physics processes and range of center-of-mass energies. \SID is based on the paradigm of particle flow, an algorithm by which the reconstruction of both charged and neutral particles is accomplished by an optimized combination of tracking and calorimetry. The net result is a significantly more precise jet
energy measurement which results in a di-jet mass resolution good enough to distinguish
between hadronically decaying $W$s and $Z$s.

\begin{figure}[tb]
 \begin{center}
 \includegraphics[width=0.9\hsize]{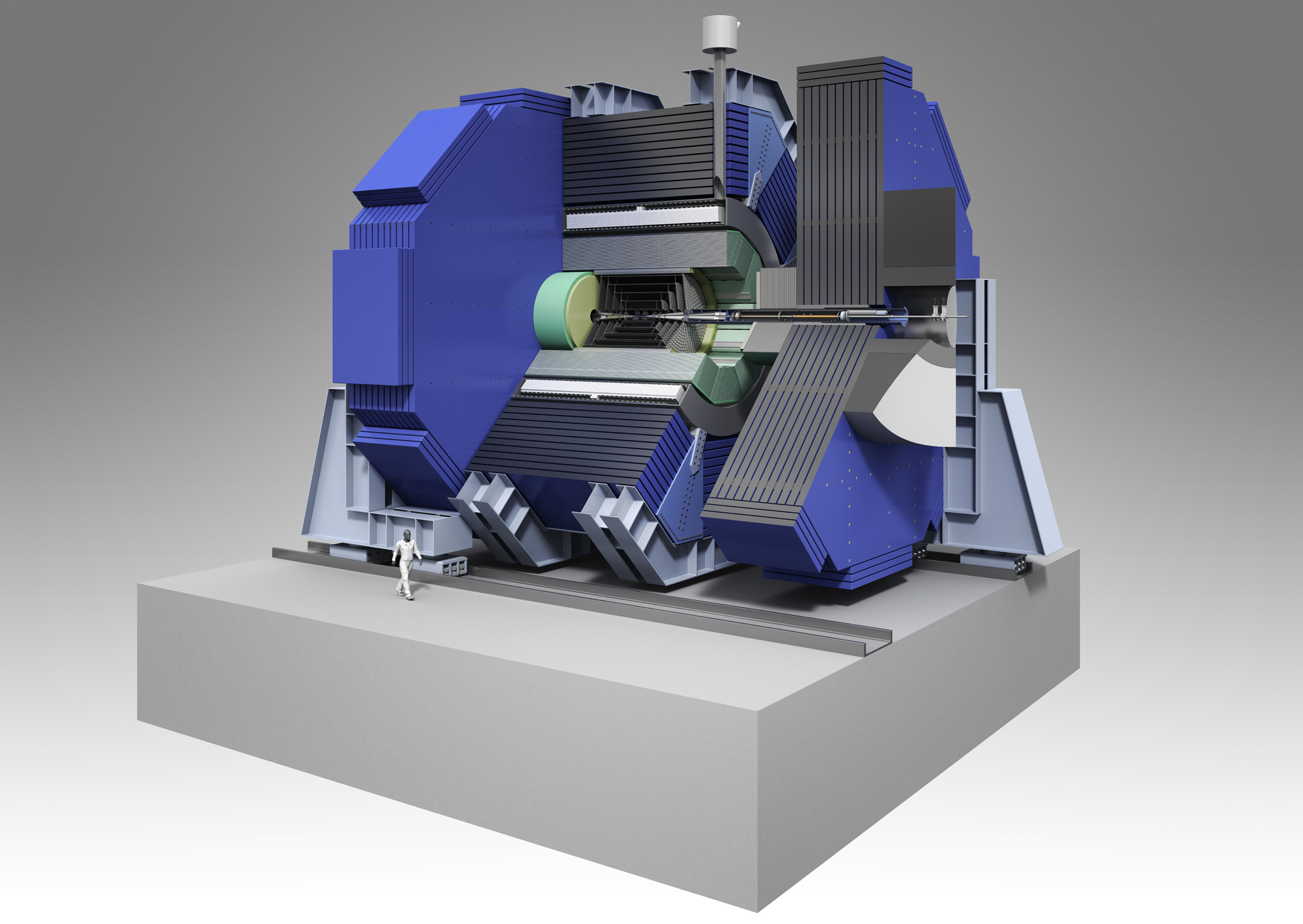}
\caption{The \SID detector as shown in the  DBD.
\label{fig:fig_sid}}
 \end{center}
 \vspace{-0.7cm}
 \end{figure}

The \SID detector (Fig.~\ref{fig:fig_sid}) is a compact detector based on a powerful silicon pixel vertex detector, silicon tracking, 
silicon-tungsten electromagnetic calorimetry, and highly segmented hadronic calorimetry. \SID also incorporates a high-field solenoid, iron flux return, a muon identification system, and forward calorimetry. 

The choice of silicon detectors for tracking and vertexing ensures that \SID is robust with respect to beam backgrounds or beam loss, provides superior charged particle momentum resolution, and eliminates out-of-time tracks and backgrounds. The main tracking detector and calorimeters can time-stamp each individual bunch crossing, so beam-related backgrounds and low-\pT background  events originating 
from $\gamma\gamma$ processes will be reduced to the minimum possible levels. 
The use of silicon sensors in the enables a unique integrated tracking system ideally suited to particle flow. The \SID calorimetry is optimized for excellent jet energy measurement using the particle flow technique. The complete tracking and calorimeter systems are contained within a superconducting solenoid, which has a \SI{5}{\tesla} field strength, enabling the overall compact design. The coil is located within a layered iron structure
that returns the magnetic flux and is instrumented to allow the identification of muons. 
All aspects of \SID are the result of intensive R\&D aimed at achieving performance at unprecedented levels. 

At the same time, the design represents a balance between cost and physics performance. The key parameters of the \SID design are listed in  
Table~\ref{sid:ConceptOverview:Table:Ovw_sidparams}.

\begin{table}[htbp]
\renewcommand{\arraystretch}{1.25}

\caption{\label{sid:ConceptOverview:Table:Ovw_sidparams}Key parameters of the baseline \SID design. (All dimension
are given in cm).}

{
\begin{tabular}{l l r r r}
    \toprule
    \SID Barrel& Technology& In rad& Out rad& z extent \\
    \midrule
    Vtx detector& Silicon pixels& 1.4& 6.0& $\pm \quad 6.25$ \\
    Tracker& Silicon strips& 21.7& 122.1& $\pm \quad 152.2$ \\
    ECAL& Silicon pixels-W& 126.5& 140.9& $\pm \quad 176.5$ \\
    HCAL& RPC-steel& 141.7& 249.3& $\pm \quad 301.8$ \\
    Solenoid& 5 Tesla SC & 259.1& 339.2& $\pm \quad 298.3$ \\
    Flux return& Scint-steel& 340.2 & 604.2& $\pm \quad 303.3$ \\
    \bottomrule

   \toprule
 \SID Endcap& Technology& In z& Out z& Out rad \\
    \midrule
Vtx detector& Silicon pixels& 7.3& 83.4& 16.6 \\
Tracker& Silicon strips& 77.0& 164.3& 125.5 \\
ECAL& Silicon pixel-W& 165.7& 180.0& 125.0 \\
HCAL& RPC-steel& 180.5& 302.8& 140.2 \\
Flux return& Scint/steel& 303.3& 567.3& 604.2 \\
LumiCal& Silicon-W& 155.7& 170.0& 20.0 \\
BeamCal& Semicond-W& 277.5& 300.7& 13.5 \\
    \bottomrule
\end{tabular}
}

\end{table}

\section{Changes to the Baseline post-DBD}
With the completion of the DBD and the intention from the Japanese HEP community 
to host the ILC in Japan, two major design changes were made to the baseline 
designs, that was presented in the DBD, the switch from a purely digital hadron 
calorimeter (DHCAL) with RPCs as active 
medium~\cite{1748-0221-5-02-P02007,1748-0221-4-10-P10008, 
1748-0221-4-06-P06003,1748-0221-4-04-P04006} to a scintillator-based solution 
with analog read-out (AHCAL) and the change of the iron yoke from a octagon to a 
dodecagon. The first choice was driven by the huge progress in the SiPM 
technology in terms of noise and stability, while at the same time the 
limitations of a large-scale RPC system with several million individual channels 
in terms of uniformity, calibration and long-term stability became more clear. 
As there were neither clear cost or performance benefits of the DHCAL at the 
time, this led to a switch to an AHCAL solution. The mechanical structure of the 
HCAL was left unchanged.

From a systems point of view, the elimination of both a \SI{7}{\kilo\volt} HV system 
and an elaborated gas system had of course significant implications, as no other other sub-detector of \SID required such systems.

The second choice was mainly driven by site-specific studies for the potential Kitakami site. 
The \SID iron yoke is assembled by stacking of eleven individual iron plates into wedges. 
By switching the iron-yoke geometry from an octagon to a dodecagon, the weight of the individual 
plates could be kept below \SI{30}{\tonne}, allowing easier transport by truck on Japanese highways.  
At the same time the overall yoke design changed from a vertical interface between
the barrel and endcap to a \SI{30}{\degree} interface. 

\begin{figure}[tb]
 \begin{center}
 \includegraphics[width=0.9\hsize]{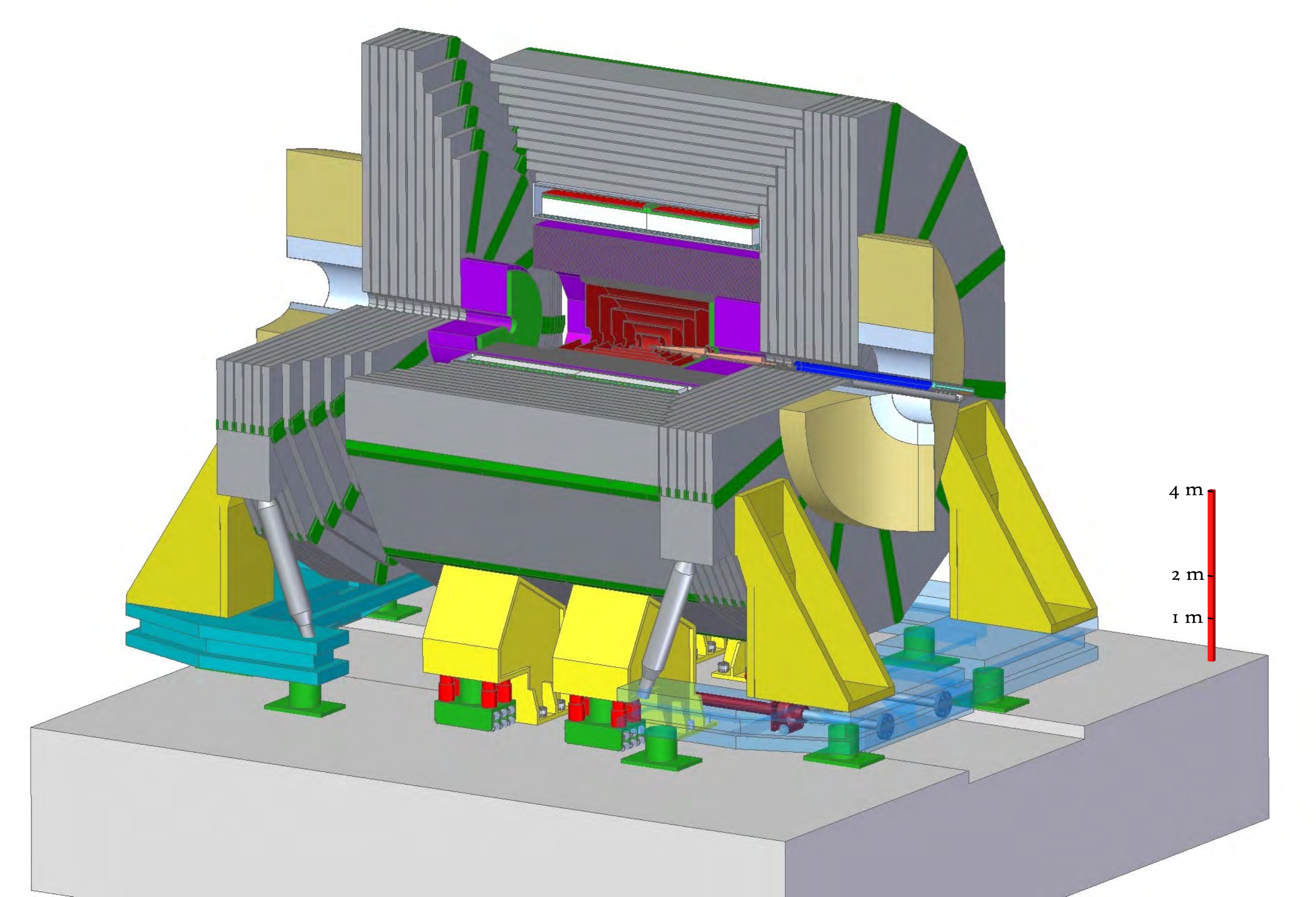}
\caption{The \SID detector concept with the reconfigured dodecagonal iron yoke.
\label{fig:fig_sidpostdbd}}
 \end{center}
 \vspace{-0.7cm}
 \end{figure}

\section{Updating the \SID detector design}
The last time the \SID detector design received a major overhaul was in the preparation of the DBD. 
With almost a decade past, technology choices, in particular, for the individual subsystems need to be reviewed. 
In the follow section, potential updates are outlined showcasing also the areas where dedicated R\&D is needed and new contribution would be extremely welcome.
Also with the advances in technology driven e.g. by the HL-LHC in terms of silicon sensor and ASIC development, 
novel timing detectors and improved services for power distribution and data transmission, it is very interesting to explore how these could be 
incorporated into an updated \SID detector concept. 

The use of Particle Flow Algorithms (PFA) for reconstruction puts significant 
constraints on the detector design. PFA detectors meed to have both the tracker 
and the calorimeter inside the solenoid for the best possible performance. 
A key ingredient of any PFA detector is a combination of large tracker radius and a strong field to separate the clusters from charged and neutral particles. Secondly a very high granularity helps to separate the individual clusters and also is important to  minimize the confusion, i.e. 
assigning individual calorimeter hits to the wrong cluster. 
The most commonly used PFA algorithm is PandoraPFA~\cite{pandorapfa,Marshall:2015rfa}, 
which has also been used extensively by \SID.

\subsection{Overall detector geometry}
The overall detector geometry has been quite stable, but there have been a few areas of discussion over the last year, which need to be revisited on the road toward a TDR for \SID.
Particularly the tracker geometry and the calorimeter thickness have a huge impact on both physics performance and cost. 
In the following a few points to be revisited and optimized are  described in more detail.

\subsubsection{Tracker radius and aspect ratio}
The outer radius of the tracker was fixed at \SI{1.25}{\meter}, mainly driven by the need to fit both tracker and calorimeter inside the solenoid and the maximum radius of a solenoid with a \SI{5}{\tesla} field. Having a larger radius would help in the shower separation, which is beneficial for the PFA reconstruction. However, the current layout is already quite close to the maximum coil radius, so the impact of a modest increase of the radius needs to be studied. It was already concluded for the DBD, that going significantly below \SI{1.25}{\meter} radius, severely impacts the PFA performance~\cite{Stanitzki:2009mx}.  For the DBD  the tracker length was fixed to \SI{331.4}{\centi\meter}, as the  tracker mechanical support structure at the time would not allow a significant extension in length. Simulation studies have shown~\cite{Stanitzki:2009mx}, that moving the transition between barrel and end-cap further out in $z$ could be beneficial for the overall performance of \SID. 

\subsubsection{Thickness of the calorimeter system}
The \SID calorimeter thickness (20+10 layers in the ECAL, 40 layers in the HCAL) was optimized to provide an excellent jet resolution from $\sqrt{s}=$91~GeV to 1~TeV. However the number of layers and total thickness of the system could be re-optimized, as the calorimeter system is a major cost driver. Studies for the ECAL as the major cost-driver have already shown, that for pure electromagnetic energy performance a thinner calorimeter would be sufficient~\cite{Braun:2020ivn}.

\subsection{Monolithic Active Pixel Sensors}
The \SID silicon baseline as described in the DBD based on two large area 
high-resistivity silicon diode arrays for both the main tracker and the ECAL 
that are read out with the bump bonded \KPIX. The possibility of stitching 
reticules together with deep implants~\cite{Ballin:2011jq} and depletion of the 
epitaxial layer make Monolithic Active Pixel sensors (MAPS) extremely 
attractive. \SID has currently focused its efforts on the Tower-Jazz 
\SI{65}{\nano\meter} process~\cite{napa}. The fundamental specifications are 
sensors of at least \SI[product-units = power]{5 x 20}{\centi\meter} with pixels 
of \SI[product-units = power]{25 x 50}{\micro\meter}. The same MAPS sensor could 
be used for the tracker and the ECAL. The sensor is readout by digital 
electronics on a "balcony" \SI{0.5}{\milli\meter} high on one edge of the 
sensor. Each pixel should have at least  simple discriminator (one bit) and 
needs to records the time for bunch identification. Distribution of power over 
such a large area is challenging. Power over Fiber with energy storage in 
capacitors near each reticle is being studied, as is data transmission by fiber.

\subsection{Vertex Detector}
Since the DBD, \SID uses a vertex detector design with five individual layers in the barrel and four disks in the endcaps. Taking advantage of the large \SI{5}{\tesla} field, the 
inner radius of the Vertex Detector is merely \SI{1.4}{\centi\meter} away from the interaction point. This layout has been shown to give an excellent performance, however a further refinement of the layout is needed.
The resolution requirements for the \SID Vertex Detector have remained unchanged since the DBD, a point resolution of better than \SI{3}{\micro\meter}, which 
mandates either pixels with a pure binary readout with a pixel pitch of \SI{10}{\micro\meter}, or a analog readout with a pixel pitch of 15-\SI{20}{\micro\meter}. 
While there have been many promising developments already~\cite{Mager:2016yvj,Greiner:2011zz,Sinev:2015iwr,Pernegger:2017trh}, a MAPS which is meeting all the 
\SID requirements is not yet available. Given the  currently foreseen assembly scheme, the Vertex Detector would be installed last and technology decisions can be taken much later than for other subsystems, taking full advantage of the latest technology developments.
The beam background levels at the several proposed ILC operating energies are quite different~\cite{Schutz:2018rox}, so the Vertex Detector is foreseen 
to be upgradable in a shutdown, which also opens the possibility to design an optimal detector for each center-of-mass energy.

The impact of the layout and resolution as well as its material budget of the Vertex Detector on the flavor tagging performance need to be revisited. 
These studies will then also guide the design of a dedicated MAPS for the Vertex Detector.

\subsection{Tracker}

The Tracker design described in the DBD is based on high-resistivity Silicon-strip sensors with a size of 10 $\times$ \SI{10}{\centi\meter} and a strip-pitch of \SI{25}{\micro\meter} 
readout by a System on Chip (SOC) \KPIX~\cite{6551433}, bump bonded to the Silicon sensor. 
The sensor uses a second metal layer to connect the individual strips to \KPIX, and no hybrid is required. A fully functional prototype has recently been 
assembled, tested~\cite{Brau:2020zrg} and achieved the desired resolution of \SI{7}{\micro\meter}. The overall tracker layout uses  five nested cylinders in 
the barrel region and four disks following a conical surface with an angle of 5 degrees with respect to the normal to the beamline in each of the endcaps. In the 
endcaps, the disks use two sensors mounted back to back to allow a small-angle stereo configuration. 

A MAPS-based tracker for \SID would have sensor sizes optimized for \SI{300}{\milli\meter} wafers. The sensors would be aligned with the \SI{25}{\micro\meter} 
pixel dimension in the bend direction, and would have a resolution of 25/$\sqrt{12}$ as charge sharing is not assumed.  
For the endcaps, such a sensor would eliminate the need for two sensors in a small-angle-stereo configuration, reducing both the material budget and cost.

\subsection{Common Tracking aspects}
For \SID, the vertex detector and main tracker have always been treated as an integrated system. For the overall system the 
question remains if the current configuration could be improved and while the tracking resilience against backgrounds 
has been extensively tested, a detailed study of the robustness of the tracker towards failures of individual channels or 
complete sensors would be further input to the overall layout.
Alignment of the tracking system will remain a challenge, as the level of e.g. Z$\rightarrow\mathrm{\mu^+\mu^-}$ events 
will be too small for beam-based alignment. Integrated scanning interferometers are expected to have adequate precision.
With the advent of fast timing and 4D-tracking, the impact on the performance of the \SID tracking system needs to be demonstrated.

Future studies for the tracking system include 
\begin{itemize}
\item Optimal Pixel size for the Vertex Detector and the tracker
\item Mechanical support structures, services and alignment studies
\end{itemize}

\subsection{Electromagnetic Calorimeter}
The ECAL is a sampling calorimeter using tungsten as a radiator with $\approx$ \SI{1}{\milli\meter} gaps for the Silicon sensors. 
The TDR design~\cite{Behnke:2013lya} employs high resistivity sensors cut into hexagons that 
are as large as practical from a six inch wafer. Similarly to the Tracker, a \KPIX ASIC is bump-bonded to each sensor. 
This concept has been successfully demonstrated in several beam tests~\cite{Barkeloo:2019zow}
Given the huge technical progress made for MAPS, a study has started to consider large- area MAPS for the ECAL, eliminating 
the need for the delicate and expensive bump-bonding, while significantly improving the separation of showers and particle passages within the electromagnetic calorimeter. 

\begin{table}[htbp]
\renewcommand{\arraystretch}{1.25}
\caption{\label{Table:showerres}The resolution for photon showers versus energy for various reconstruction modes}
\begin{tabular}{l c}
\toprule
All MIPS & $8.8\%/\sqrt{E} \oplus 0.2\%$\\
active pixels with MIPS&   $9.8 \%/\sqrt{E} \oplus 1.1\%$\\
weighted clusters& $12.2\%/\sqrt{E} \oplus 1.4\%$\\ 
clusters &$13.7 \%/\sqrt{E} \oplus 1.9\%$ \\  
Hits &$16.4\%/\sqrt{E} \oplus 2.0\%$\\ 
TDR   &$17 \%/\sqrt{E} \oplus 1.0\%$\\\bottomrule
\end{tabular}
\end{table}

\begin{figure}[htbp]
\begin{center}
\includegraphics[width=0.9\hsize]{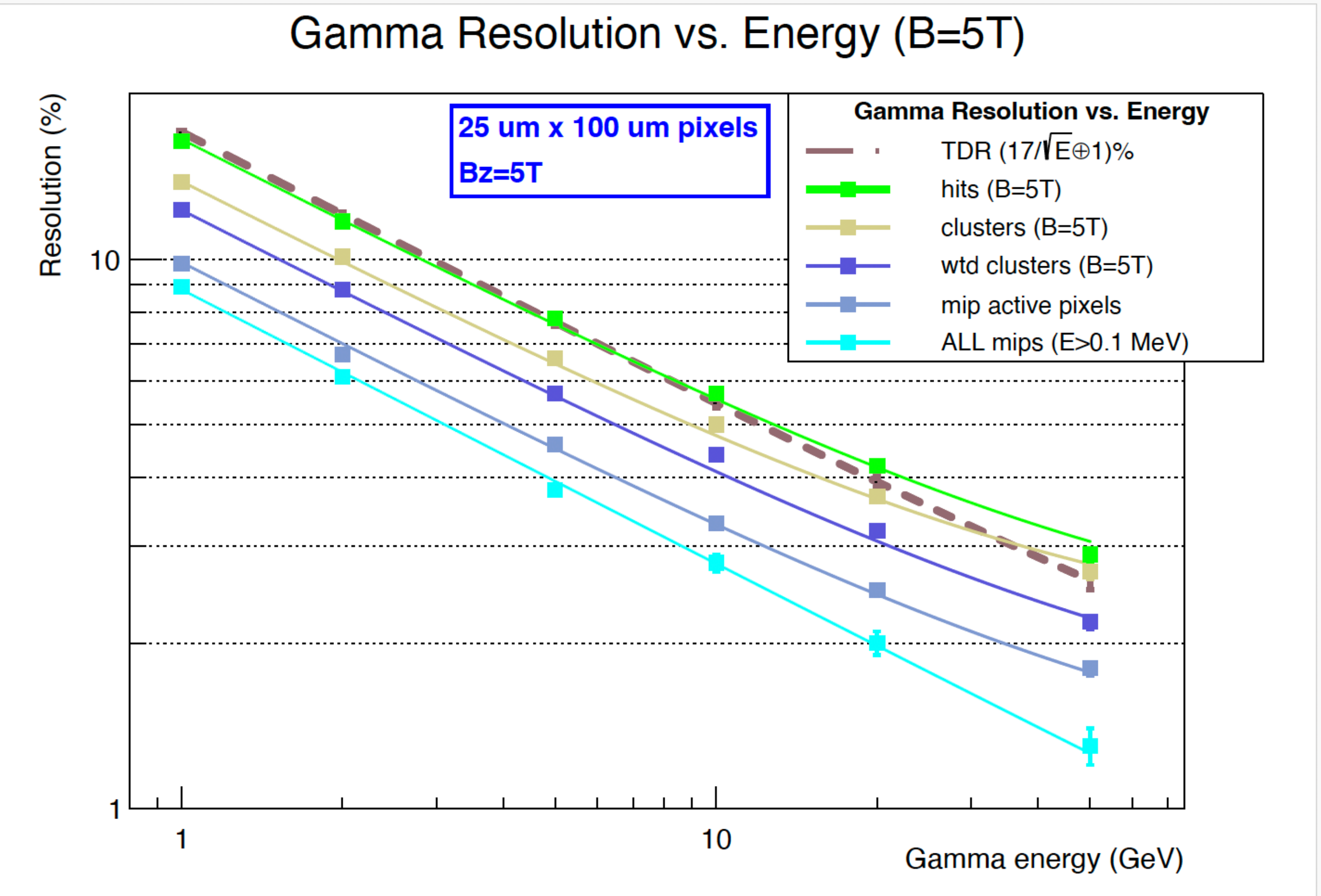}
\caption{Energy resolution for photon showers.~\cite{Brau:2024jrq,Brau:2025cal}
\label{fig:fig_New_Eres}}
 \end{center}
 \vspace{-0.7cm}
\end{figure}

As discussed before, a pixel area of \SI{2500}{\micro\meter\tothe{2}}  (\SI[product-units = power]{50 x 50}{\micro\meter} or \SI[product-units = power]
{25 x 100}{\micro\meter}) is considered to be a sensible choice, and excellent performance with a purely digital ECAL based on this 
fine granularity is expected, but the use of some analog information will be investigated.  A pixel area of \SI{625}{\micro\meter\tothe{2}}  (\SI[product-units = power]{25 x 25}{\micro\meter}) is also being considered.
Previous studies~\cite{Ballin:2009yv,Stanitzki:2011zz,Dauncey:2010zz} have even indicated potential energy resolution advantages 
for a digital ECAL solution. New simulation studies, based on this finer, digital configuration, have 
confirmed the previous studies and demonstrated additional details on the performance~\cite{Brau:2022}. These 
studies indicate the electromagnetic energy resolution based on counting 
clusters of hits in the MAPS detectors should provide better performance than 
the \SID original design based on  \SI{13}{\milli\meter\tothe{2}} analog pixels. 
In addition, weighting of clusters relative to their distance from the shower axis improves even more.
This is shown in Figure ~\ref{fig:fig_New_Eres}.  
The resolution for photon showers versus energy ~\cite{Brau:2024jrq,Brau:2025cal} is shown in Table~\ref{Table:showerres}.

\begin{figure}[htbp]
\begin{center}
\includegraphics[width=0.9\hsize]{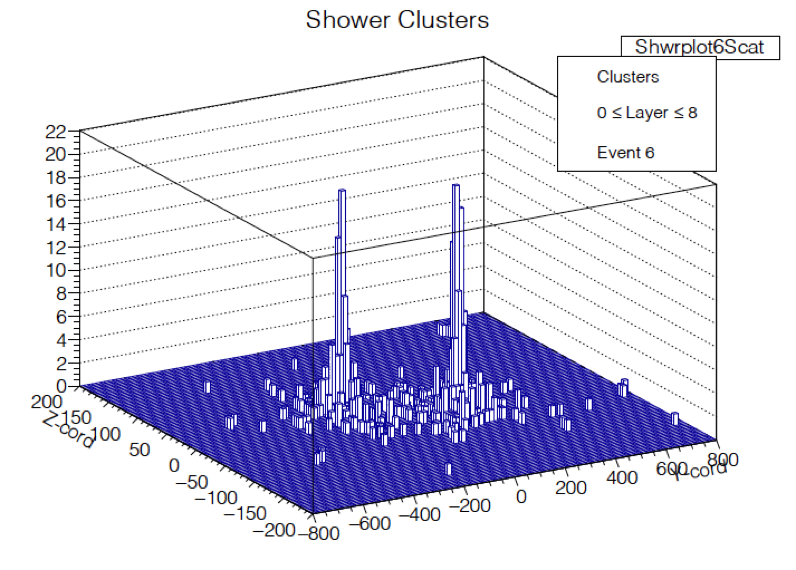}
\caption{Transverse distribution of clusters in the first 5.4 radiation lengths for two 10~GeV electron 
showers with a separation of one centimeter in the new \SID digital MAPS~\cite{Brau:2022}.
\label{fig:fig_New_Two}}
 \end{center}
 \vspace{-0.7cm}
\end{figure}

\begin{figure}[htbp]
\begin{center}
\includegraphics[width=0.9\hsize]{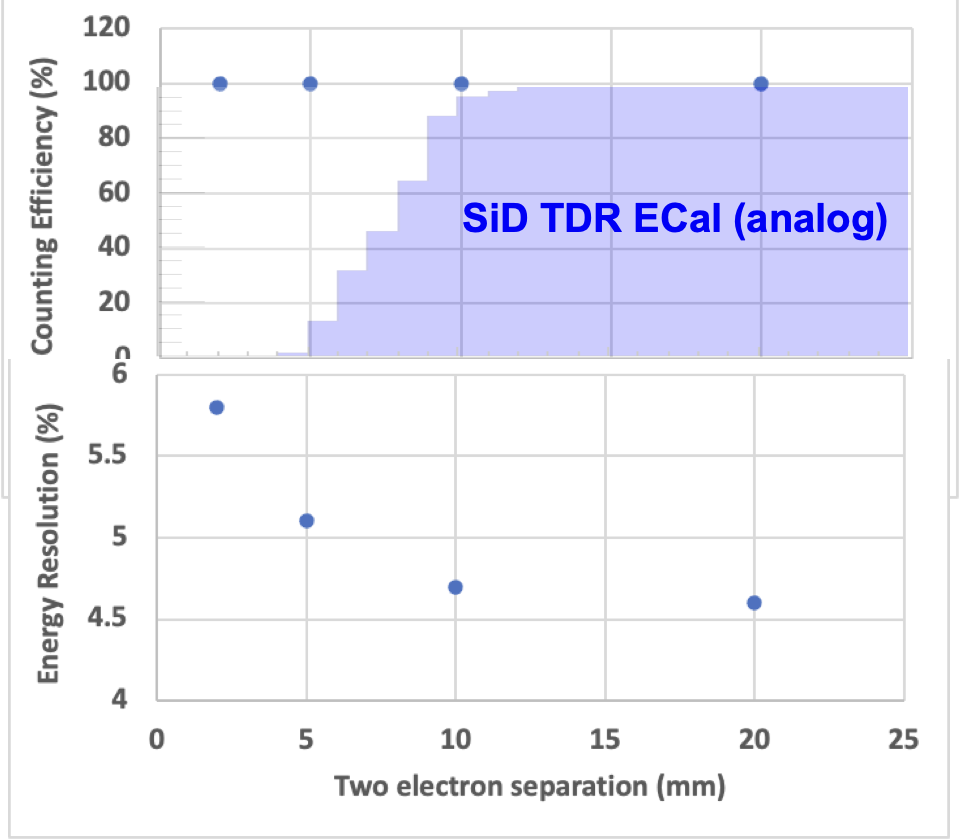}
\caption{Efficiency for distinguishing two 10~GeV electron showers as a function of shower separation (upper) and the degradation of energy resolution as a function of separation due to overlap of cluster hits (lower) in the new \SID digital MAPS ~\cite{Brau:2022}.
\label{fig:fig_New_Sum}}
 \end{center}
 \vspace{-0.7cm}
\end{figure}

Furthermore, two shower separation is excellent, 
as shown in Figure ~\ref{fig:fig_New_Two}. The performance for two showers versus their 
separation is summarized by Figure ~\ref{fig:fig_New_Sum}.  The fine granularity 
allows for identification of two showers down to the millimeter scale of 
separation, and the energy resolution of each of the showers does not degrade 
significantly for the millimeter scale of shower separation. 

\begin{figure} [htbp]
\begin{center}

\includegraphics[width=0.95\hsize]{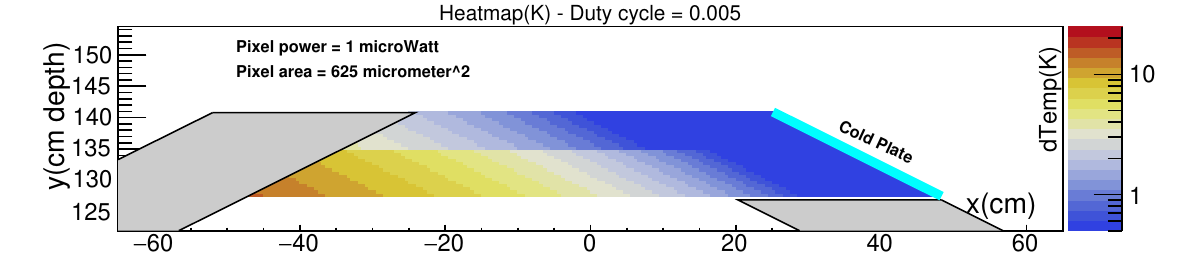}
\caption{Model of the heat map of the SiD ECAL barrel wedge operating with the ILC duty cycle and using pixel sizes of  \SI[product-units = power]{25 x 25}{\micro\meter}.
For the lower duty cycle of \CCC or CLIC the heat rise is even smaller.}
\label{fig:heat}       
\end{center}  
\vspace{-0.7cm}
\end{figure}

In order to preserve the \SI{14}{\milli\meter} Moliere radius of the calorimeter with the \SI{1.25}{\milli\meter} readout gap, power pulsing and passive cooling is used.  For this, the power pulsing enabled by the linear collider time structure is essential. 
One end of each ECAL wedge provides a cold plate and the heat is conducted from the sampling layers through the tungsten layers to the cold plate.  
For the sparsest time structure of \CCC this leads to a maximum temperature rise of 2~K at the far corner of the module, and a rise of 16~K for ILC operation which has beam bunching with a higher duty cycle of 0.5\% ~\cite{Brau:2024jrq,Brau:2025cal}.
Figure \ref{fig:heat} presents an estimate of the temperature rise throughout the barrel module for the ILC bunch train.

Future studies include: 
\begin{itemize}
\item Optimization of pixel size for the digital option and investigating potential for multi-bit digital operation;
\item Complete mechanical design of modules including verification of passive heat removal;
\item Optimization of the overall ECAL design including manufacturabilty;
\end{itemize}

\subsection{Hadronic calorimeter}
\begin{figure}[htbp]
 \begin{center}
 \includegraphics[width=0.9\hsize]{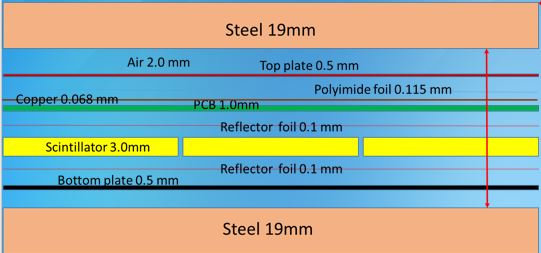}
\caption{HCAL simulated layer design.
\label{fig:fig_HCAL_layer}}
 \end{center}
 \vspace{-0.7cm}
 \end{figure}

The current baseline design for the HCAL has active layers of small ($\SI{3}{\centi\meter} \times \SI{3}{\centi\meter}$) scintillator tiles read out by silicon photomultipliers (SiPM), between steel absorber layers. 
There are 40 layers in the alternating active/absorber structure. There has been considerable development of this technology by the CALICE 
collaboration ~\cite{Sefkow:2015hna}. The \SID simulation  includes a detailed description of the CALICE active layer design 
(Fig.~\ref{fig:fig_HCAL_layer}).

\begin{figure}[htbp]
 \begin{center}
 \includegraphics[width=0.9\hsize]{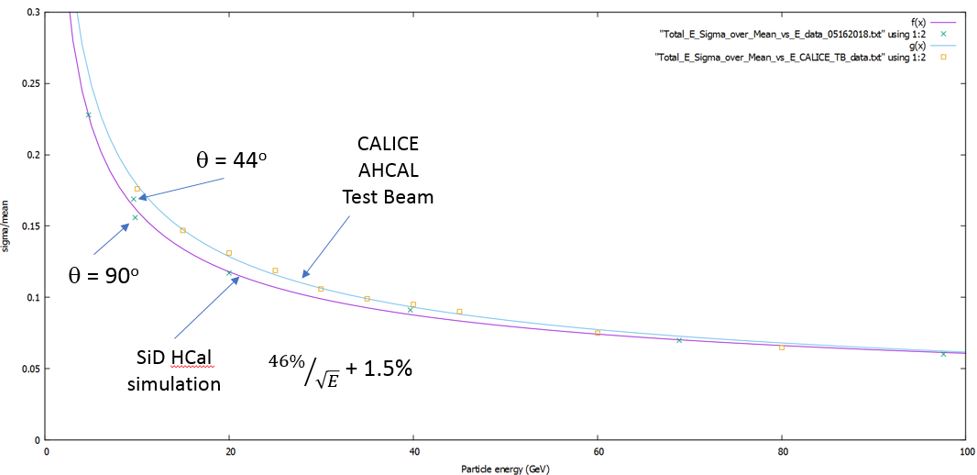}
\caption{HCAL energy resolution - simulated and prototype results.
\label{fig:fig_HCAL_energy_resolution}}
 \end{center}
 \vspace{-0.7cm}
 \end{figure}

To verify the performance of the \SID HCAL simulation, the single pion energy resolution has been compared with the results 
from the large-scale CALICE prototype. The comparison is shown in (Fig.~\ref{fig:fig_HCAL_energy_resolution}) and shows good agreement between the simulation and the CALICE test beam results~\cite{1748-0221-10-04-P04014}.

 Previous simulation results have also shown that jet energy resolution in the 3-4\% range can be achieved. However, as yet, the effects of variations and uncertainties arising from calorimeter design such as the number of layers, ratio of active to absorber layer thicknesses, calibrations within and between layers, transition regions between electromagnetic and hadronic systems and barrel and endcaps, and those deriving from e.g. confusion in the particle flow algorithm, have yet to be considered in terms of their influence on precision physics results. Therefore, detailed studies of these effects and their potential for improving the HCAL design are needed. 
 Extensions and optimization of the design are being discussed with the DRD6 collaboration and include the following:
 \begin{itemize}
\item Inclusion of timing layers to assist the particle flow algorithm in
separating the delayed shower components from slow neutrons from the prompt
components. 
\item Inclusion of precision tracking layers to assist the PFA.

\item Exploration of benefits of using on-board intelligence, ranging from
simple zero suppression to interlayer communication to assist tracking through
the HCAL and or PFA jet reconstruction.

\item  The transition from very fine MAPS ECAL cells to larger HCAL cells should be studied to optimize performance.
 
\item Potential cost savings by making some of the outer layers thicker if there
is no significant degradation in energy resolution. 

\item Reconsideration of the effects of projective cracks between modules. There 
is some indication from earlier studies that projective cracks have no negative 
effect on energy resolution, but this needs further verification.

\item Exploration of alternative layouts for HCAL sectors in the end-caps and optimization of the boundary between the barrel and end-caps.

\end{itemize}

\subsection{Superconducting Coil}
The \SID solenoid is based on the CMS solenoid
design philosophy and construction techniques, using a slightly modified CMS
conductor as its baseline design. Superconducting strand count in the coextruded
Rutherford cable was increased from 32 to 40 to accommodate the higher \SI{5}{\tesla}
central field. 

Many iron flux return configurations have been simulated in two
dimensions so as to reduce the fringe field. An Opera 3D calculation with the Detector
Integrated Dipole (DID) coil has been completed and magnetic field calculations with ANSYS are in progress. These will have the capability to calculate forces and stress on the DID as well as run transient cases to check the viability of using the DID as a quench propagator for the solenoid. Field and force calculations with
an iron endcap HCAL were studied. The field homogeneity improvement was found
to be insufficient to pursue this option. 

Conceptual DID construction and assembly methods have been studied. The solenoid 
electrical power system,including a water-cooled dump resistor and grounding, 
was established. Significant work has been expended on examining different conductor stabilizer
options and conductor fabrication methods. This work is pursued as a cost- and
time-saving effort for solenoid construction.

Preliminary work has begun on the possibility of replacing the CMS conductor 
with Cable in Conduit Conductors (CICC). These cables are conduits of stainless 
steel with Ni-Ti conductor (for our relatively low fields) inside. The cable is 
cooled by flowing either superfluid or supercritical He. The advantages include 
elimination of a separate He vessel, smaller conductor cross section because of 
the tensile strength advantages of stainless over Aluminum, and easier 
construction. The interesting possibility is that the solenoid package thickness 
could be substantially reduced, leading to large cost reductions.

\subsection{Muon Detectors}

The flux-return yoke is instrumented with position sensitive detectors to
serve as both a muon filter and a tail catcher. The total area to be
instrumented is very significant -- several thousand square meters. Technologies
that lend themselves to low-cost large-area detectors are therefore under
investigation. Particles arriving at the muon system have seen large amounts of
material in the calorimeters and encounter significant multiple scattering
inside the iron. Spatial resolution of a few centimeters is therefore
sufficient. Occupancies are low, so strip detectors are possible. The \SID
baseline design uses uses extruded scintillator readout with wavelength 
shifting fibers and SiPMs, which has been successfully demonstrated ~\cite{Denisov:2016vgm} in terms of position and time resolutions. 
Simulation studies have indicated that nine or more layers of sensitive detectors 
yield adequate energy measurements and good muon detection efficiency and purity.
The flux-return yoke itself has been optimized with respect to the
uniformity of the central solenoidal field, the external fringe field,
and ease of the iron assembly. This was achieved by separating the  barrel and end sections of the
yoke along a \SI{30}{\degree} line.

Further development of the muon system require:

\begin{itemize}
\item Optimization of layout, design and number of instrumented layers for both  barrel and end-caps.
\item Occupancies at inner radius of muon end-caps versus strip widths.
\item Role of the muon system as tail-catcher and consideration and implications of the CALICE ECAL + HCAL + Tail-catcher results.
\item Potential for use of muon system in search for long-lived particles.
 \end{itemize}

\section{Forward systems}
Two special calorimeters are foreseen in the very forward region: LumiCal for a 
precise luminosity measurement, and BeamCal for the fast estimation of the 
collision parameters and tagging of forward-scattered beam particles. LumiCal 
and BeamCal are both compact cylindrical electromagnetic calorimeters centered 
on the outgoing beam, making use of semiconductor-tungsten technology. BeamCal 
is placed just in front of the final focus quadrupole and LumiCal is aligned 
with the electromagnetic calorimeter endcap. 
LumiCal makes use of conventional silicon diode sensor readout. It is a 
precision device with challenging requirements on the mechanics and position 
control, and must achieve a small Moliere radius to reach its precision targets. 
Substantial work has been done to thin the silicon sensor readout planes within 
the silicon-tungsten assembly. Dedicated electronics with an appropriately large 
dynamic range is under development.
BeamCal is exposed to a large flux of low-energy electron-positron pairs 
originating from beamstrahlung. These depositions, useful for a bunch-by-bunch 
luminosity estimate and the determination of beam parameters, require radiation 
hard sensors. The BeamCal has to cope with 100\% occupancies, requiring 
dedicated front-end electronics. A challenge for BeamCal is to identify sensors 
that will tolerate over one MGy of ionizing radiation per year. Sensor 
technologies under consideration include polycrystalline chemical vapor 
deposition (CVD) diamond (too expensive to be used for the full coverage), GaAs, 
SiC, Sapphire, and conventional silicon diode sensors. The radiation tolerance 
of all of these sensor technologies has been studied in a high-intensity 
electron beams. 

Tasks remaining for the forward calorimeters, with participation in the FCAL R\&D 
Collaboration, include:

\begin{itemize}
\item LumiCal: complete development of large dynamic range readout electronics.

\item LumiCal: develop and demonstrate the ability to position and maintain the position of the calorimeter, particularly at the inner 
radius, in view of the steep dependence of the rate of Bhabha events on polar angle.

\item BeamCal: continue the search for and testing of suitable sensor technology(s) capable of sustained performance.
\item BeamCal: continue the study of recognizing single electron shower patterns for tagging for physics studies in the face of high radiation background.
\end{itemize}

\section{Electronics \& DAQ}
The Electronics and DAQ  needs are strongly tied to the  linear collider environment, with short bunch trains ( a few microseconds to \SI{1}{\milli\second}) 
followed by a long quiet time (~10 ms to \SI{199}{\milli\second}), 
enabling front-end buffering and read-out during the quiet time and with power-pulsing of the front-ends to reduce the overall power consumption. At the same time the physics event rate is 
a lot smaller than at e.g. the LHC and the main occupancy drivers are the beam backgrounds and the detector noise. 
\subsection{Front-end electronics}
With the incorporation of more and more MAPS-based subdetectors, the front-end ASICs become less and less widespread, and most likely the HCAL will be the first detector 
using a classical front-end for reading out the SiPMs. At the same time, the role of data concentration will only increase, bundling the data streams from several MAPS 
units into high-speed data links and moving them off-detector. More development is needed for DC-DC converters that are in the magnetic field and deal elegantly with the pulsed front end load.
\subsection{DAQ}
\begin{figure}[htbp]
\begin{center}
\includegraphics[width=0.9\hsize]{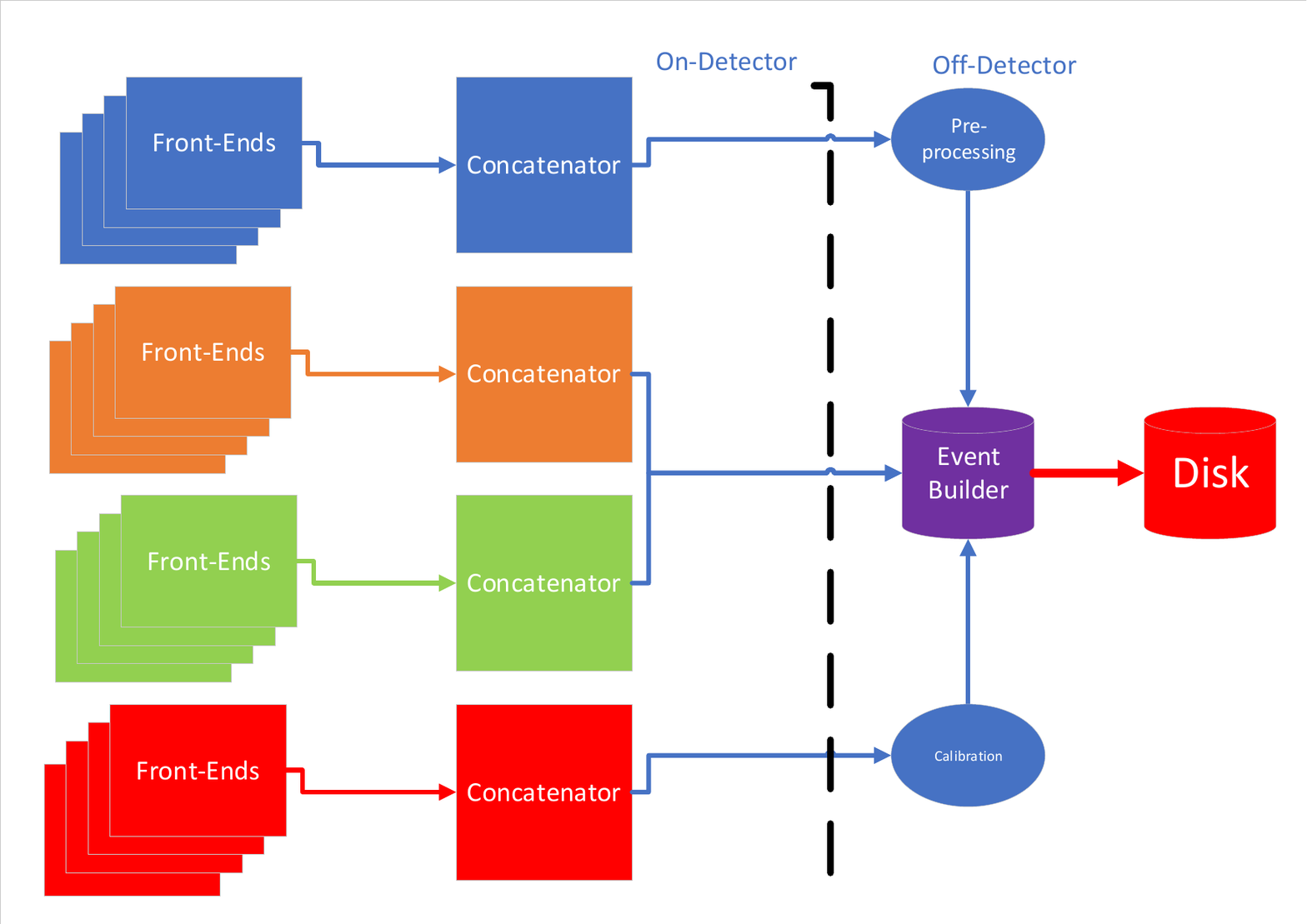}
\caption{A schematic flow in the \SID Detector with the front-end units, the data concentrators and the event building stage. \label{fig:DAQ}}
\end{center}
\vspace{-0.7cm}
\end{figure}

The data flow in any ILC detector is quite different compared to an LHC detector due the complete absence of triggering and the buffering on the front-end during the train.
As described above, at the end of a bunch train, the data of all front-ends of a sub-detector is moved to the sub-detector data concentrators
and from there, the data is sent off-detector using high-speed links. Depending on the sub-detectors, there are intermediate stages like calibration or pre-proproessing, 
before the complete events are being build by the event builder and then stored directly on disk. 
A event selection of interesting events   will be done completely offline by the physics analysis.
A sketch of the data flow is shown in Fig.~\ref{fig:DAQ}.

The DBD described a DAQ architecture based on ATCA, which was quite adequate for the relatively simple problem compared to the LHC detectors. Given the delay, the platform choice will be reviewed.

\section{Machine Detector Interface}
The Machine-Detector Interface (MDI) is a key part of any linear collider detector and essential for its performance. 
For the ILC it involves not only the machine optics like  Final-Focus magnets but more generally any interface between the detector, the machine and the interaction region hall infrastructure
Many of these activities are covered by experts from the accelerator and detector community.

\subsection{Devices within \SI{25}{\centi\meter} of the beamline}

\begin{figure}[htbp]
 \begin{center}
 \includegraphics[width=0.9\hsize]{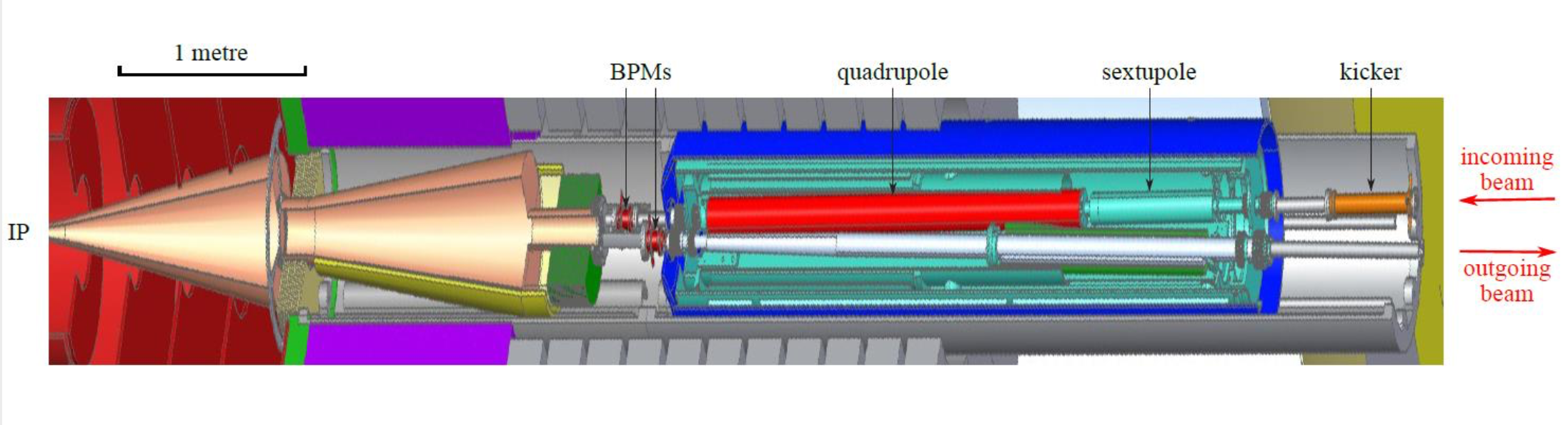}
\caption{The R25  Region of \SID. 
\label{fig:fig_MDI}}
\end{center}
\vspace{-0.7cm}
\end{figure}
 
While engineering of the R25 region of the detector has begun, much work remains. 
See for example Figure~\ref{fig:fig_MDI}.
Devices within this region include:

\begin{itemize}
 \item Beam pipe, the support system required to support the fragile beampipe between the two support tubes and Cable and Fiber pathways for power and signals
 \item Luminosity Monitor and a potential forward Hadron calorimeter
 \item Tungsten and polycarbonate masks
 \item The Feedback kicker, Feedback BPMs on the incoming and outgoing beamlines after Beamcal and before QD0
 \item The carbon fiber support tube that hold everything from the Beamcal to QD0
 \item The piezo (or other) adjustment system, resident in the return yoke, that can align the support tube, allow the \SID Endcap Door to open and functions in the magnetic environment of the powered detector.

 \end{itemize}
 
 Most of the work done in this area was completed ~2012 for the DBD. Since that time L* has changed, 
 the magnet yoke has changed and the beam parameters have changed. Optimizing the design for the final 
 configuration will require new efforts.
 
 \subsection{Devices further removed from the IP}
 Many challenging engineering and physics issues often associated with MDI need additional work. A short list includes:
 \begin{itemize}
 \item Polarimeter and energy spectrometer for both the incoming and outgoing beam as well as the gamma calorimeter
  \item PACMAN shielding to maintain self-shielding
 \item Push-Pull engineering details
 \item Anti-Detector Integrated Dipole evaluation, risk analysis and integration with the solenoid
 \end{itemize}

\subsection{Beam backgrounds}
\begin{figure}[tb]
 \begin{center}
 \includegraphics[width=0.9\hsize]{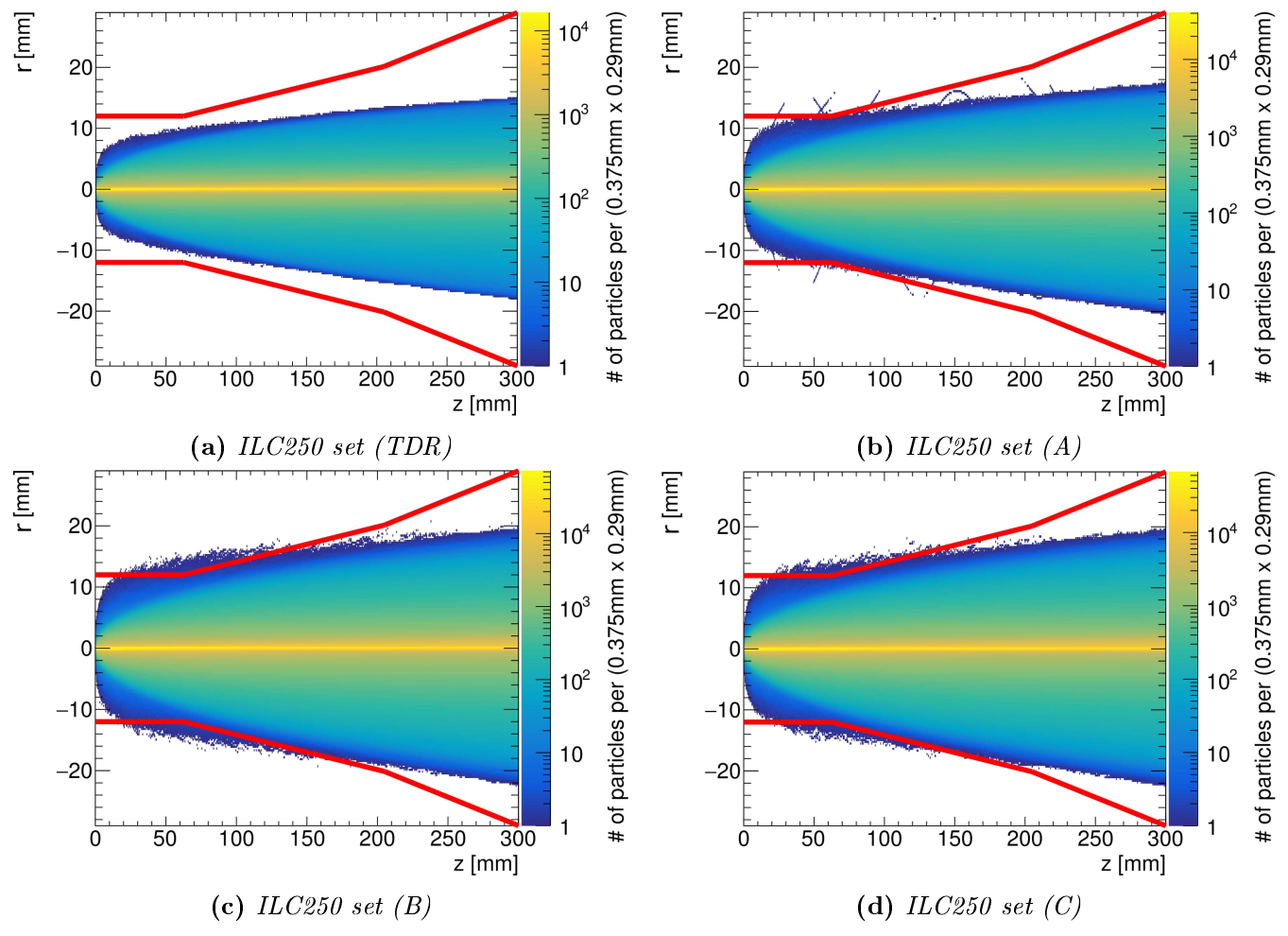}
\caption{Beam Background envelopes for several optics variants for the 250~GeV ILC~\cite{Schutz:2018ynd}. The beam pipe is indicated by the red lines. 
\label{fig:fig_beamenvelopes}}
 \end{center}
 \vspace{-0.7cm}
 \end{figure}

The main background at the ILC is due to beam-beam interaction.
The choice of machine parameters has a big impact on the \SID design, it e.g. determines the beam-pipe radius (see Figure~\ref{fig:fig_beamenvelopes}) and also 
has a big impact on the electronics design and in particular on the amount of front-end buffers necessary.

 Future studies in this area require to visit the beam background studies with the latest beam parameters and to conduct radiation damage studies for all the components close to the beam pipe.
 
\section{Software Developments}
The simulation and reconstruction software stack for \SID is based on the 
DD4HEP~\cite{1742-6596-513-2-022010} framework, which makes the simulation of 
various CALICE detectors available for evaluation in physics studies, and \SID 
uses the PandoraPFA~\cite{pandorapfa,Marshall:2015rfa} and 
LCFIPlus~\cite{Suehara:2015ura,Bailey:2009ui} packages for calorimeter 
reconstruction and flavor tagging, respectively. Looking forward, \SID is looking to keep 
pushing the envelope of modern software development.

\subsection{Programming Languages for Framework and Analysis}

\cplusplus has been the lingua franca of particle physics-related software development over the past two decades or so. However, computer science classes 
in universities are increasingly teaching other languages~\footnote{For example, \url{https://mitmath.github.io/18337/}}, 
compute accelerators like GP-GPUs are programmed in specialized languages like CUDA~\cite{10.5555/2430671}, and machine learning research is mostly done in \python and 
\julia~\cite{doi:10.1137/141000671}, and the industry is looking to \go and \rust 
for better memory safety and support of multi-threading.

\begin{figure}[htbp]

 \includegraphics[width=0.9\hsize]{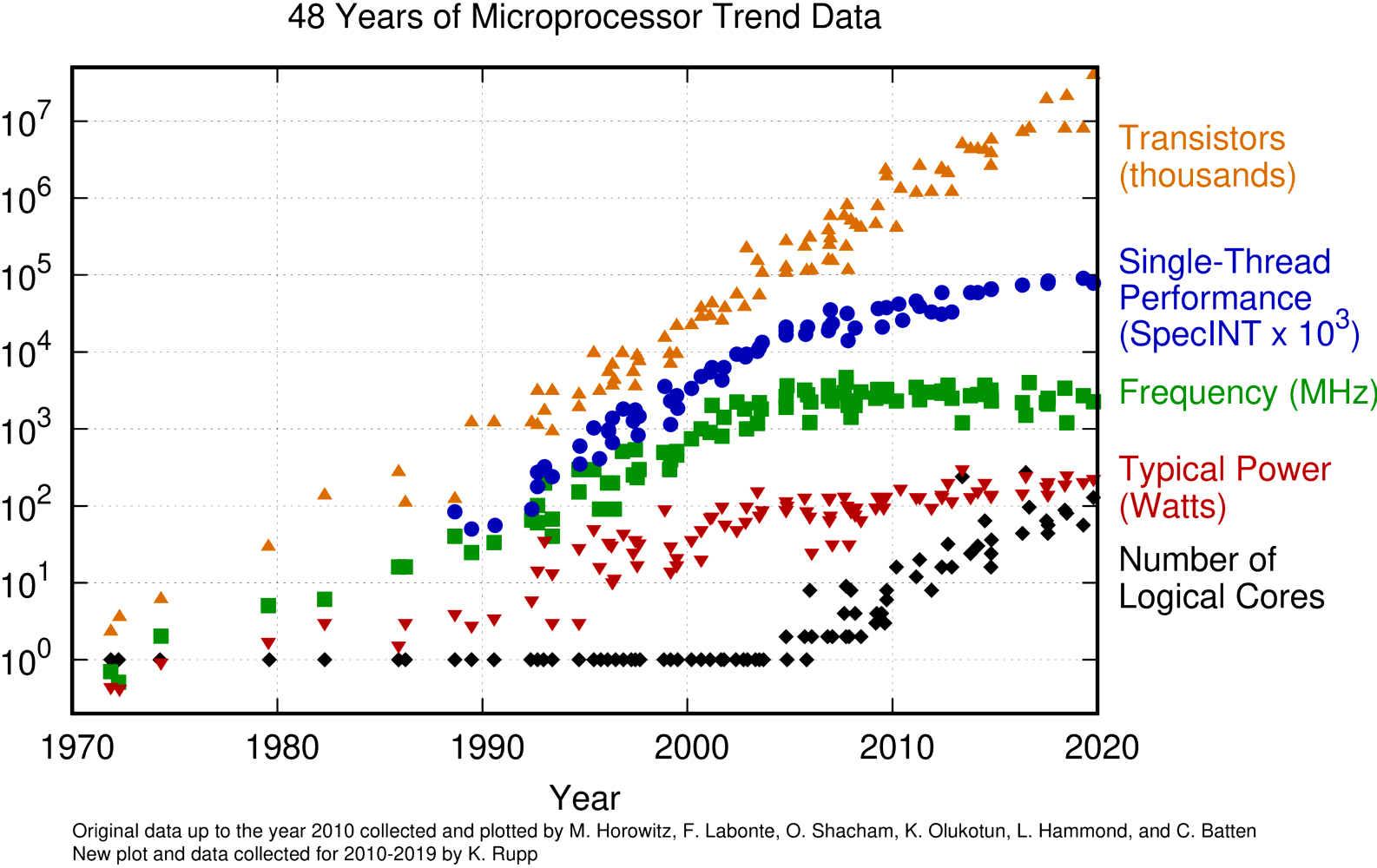}
 \caption{The growth of cores per CPU over time~\cite{microprocessor}.\label{fig:48years}}

 \end{figure}

As recent analyses~\cite{microprocessor} have shown (see Figure~\ref{fig:48years}),the single-thread performance and the closely linked clock-speed has 
been almost been leveling out since a decade, but the amount of cores per CPU has been growing exponentially. 
The overall transistor counts still follow an exponential growth line and with the the upcoming introduction of \SI{10}{\nano\meter} and \SI{7}{\nano\meter} 
processes nodes it is likely, that this trend will continue for the next few years. The gains seen for SpecINT to measure single-threaded performance are primarily 
due to compilers employing auto-vectorization and auto-parallelization~\cite{microprocessor}. Therefore, there is a clear need for a next-generation software framework 
to have the  ability to utilize multi-core and heterogeneous hardware architectures. 

In the upcoming phase of research and development we encourage the use of high-productivity languages for prototyping and design studies. 
Experience shows that many pieces of code from beam tests end up as parts of the experimental framework. The Julia programming language occupies an 
interesting point in the intersection of the high-productivity and high-performance language families~\cite{Stanitzki:2020bnx}.
This might obviate the translation of code from one language to another for performance reasons.

\subsection{Common Event Data Model(EDM)}
A common EDM has served the linear collider community very well and \lcio~\cite{Gaede:2003ip} has been very successful throughout the community. 
\lcio has bindings for many languages ranging from \fortran to \python and \julia, 
which has been another ingredient for its success. All linear collider concepts strongly supporting maintaining a common EDM, as it increases the capability to quickly cross-check results and reduces the amount of manpower needed to support a multitude of EDMs.
Updates  will support multi-threading and fast in-memory serialization to take advantage of the next-generation CPUs.

\subsection{Simulation and Reconstruction Software developments}

In the near term, \SID is going to develop an improved reconstruction for 
digital calorimetry and to study the use of timing information to improve the 
event reconstruction. Looking further ahead, the advent of machine learning 
techniques has brought many innovative solutions to jet tagging problems at the 
LHC. The level of detail available in an ILC event will lead to additional 
developments in this area. We expect reconstruction strategies using machine 
learning to have an improved performance over hand-crafted tools that take years 
to develop and often depend on only a few experts for maintenance and support. 
Reconstruction strategies using state-of-the-art machine learning methods allow 
data-driven methods for training and commissioning, and they facilitate the 
exchange of skills with other domains, which should ultimately make them easier 
to maintain and update. 

In the area of simulation, \geant~\cite{AGOSTINELLI2003250, 1610988,ALLISON2016186} is the engine of choice for \SID. However, for 
many aspects of the physics studies, this level of detail is not needed. Large 
savings in the CPU budget can be achieved by selecting a faster simulation 
procedure. Machine learning-based approaches will help to develop a faster 
simulation for large-scale production samples. Additionally, ML will help 
identify which level of detail is required for which application.

The relatively democratic distribution of processes in the electroweak sector in 
the ILC data sample might make it feasible to apply event-level matrix element 
methods to support the event interpretation. The ultimate goal of such an 
approach would be to predict the Feynman diagram based on the set of raw 
detector inputs. While this goal is quite likely out of reach for the start of 
the ILC, research in this sector can inform the detector design in the future 
and will improve the understanding of detector sensitivity to a variety of 
physics models.

\section{Potential new subsystems}
The most obvious addition to \SID is taking advantage of the recent improvements in fast-timing detectors.

\subsection{Timing layers}

As noted above the introduction of timing layers into the HCAL could allow beneficial identification of slow shower components from prompt components. 
This would provide an extra dimension to resolving track-cluster associations, and following shower development from layer to layer in the PFA. Timing resolution at the nanosecond level should be sufficient for this application.

 \begin{figure}[htpb]
 \begin{center}
 \includegraphics[width=0.9\hsize]{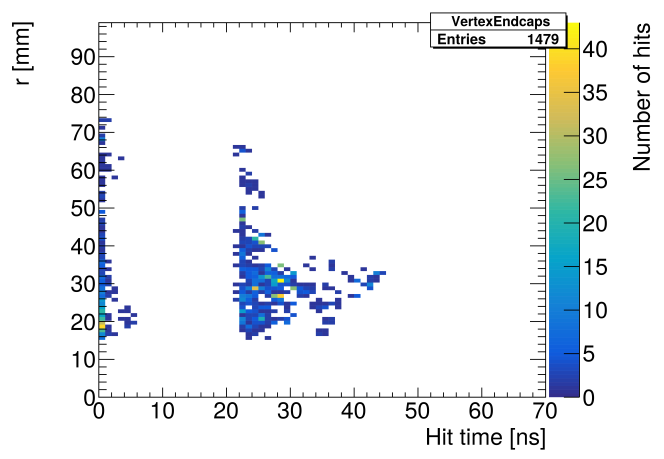}
 \caption{The time distribution of beam background hits in the \SID Vertex Detector Endcap\label{fig:vdendcaphits}}
 \end{center}
  \vspace{-0.7cm}
 \end{figure}
 
In Figure~\ref{fig:vdendcaphits} the timing distribution of the beam background hits is clearly showing the collision and then with a clear separation the backgrounds hits from a backslash from the forward region.

\begin{figure}[tb]
 \begin{center}
 \includegraphics[width=0.9\hsize]{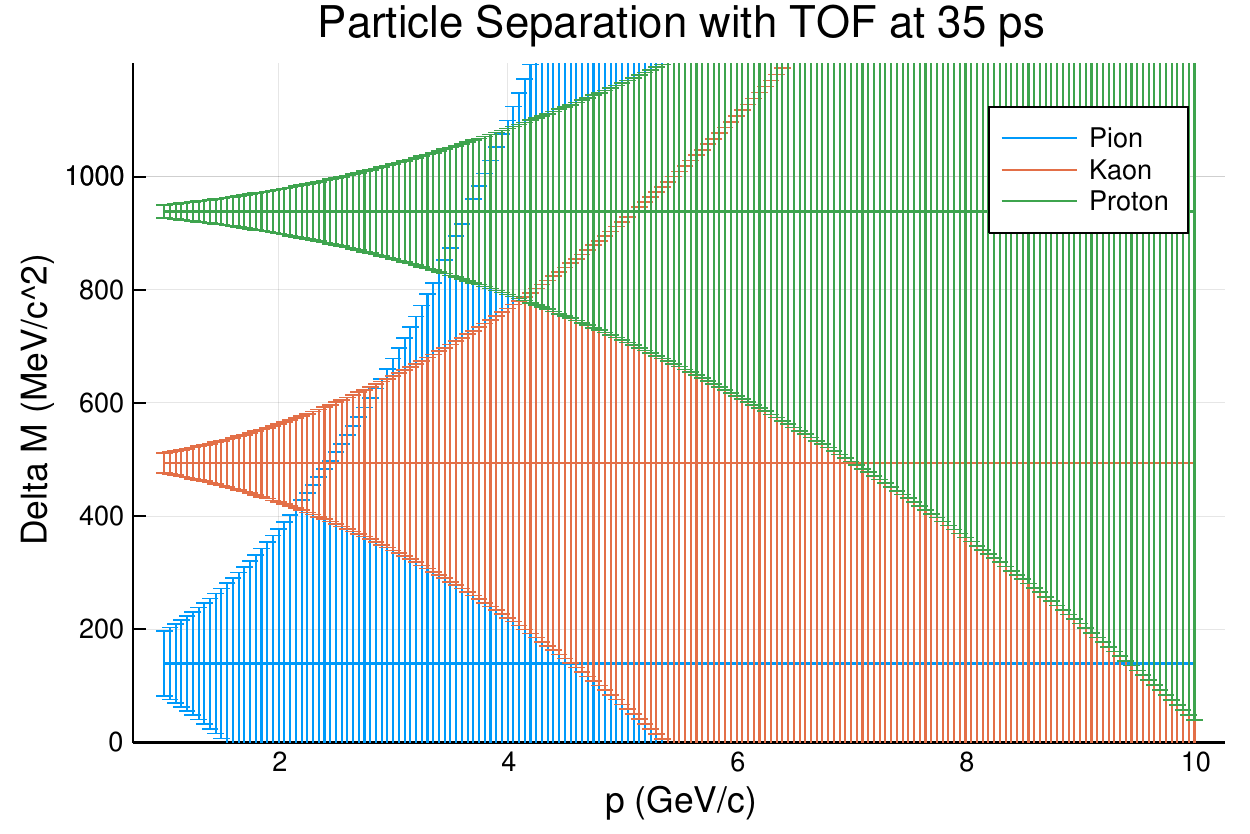}
 \caption{Mass resolution for a TOF system with a \SI{35}{\pico\second} time resolution in \SID \label{fig:Tofhllhc}}
 \end{center}
  \vspace{-0.7cm}
 \end{figure}

\begin{figure}[tb]
 \begin{center}
\includegraphics[width=0.9\hsize]{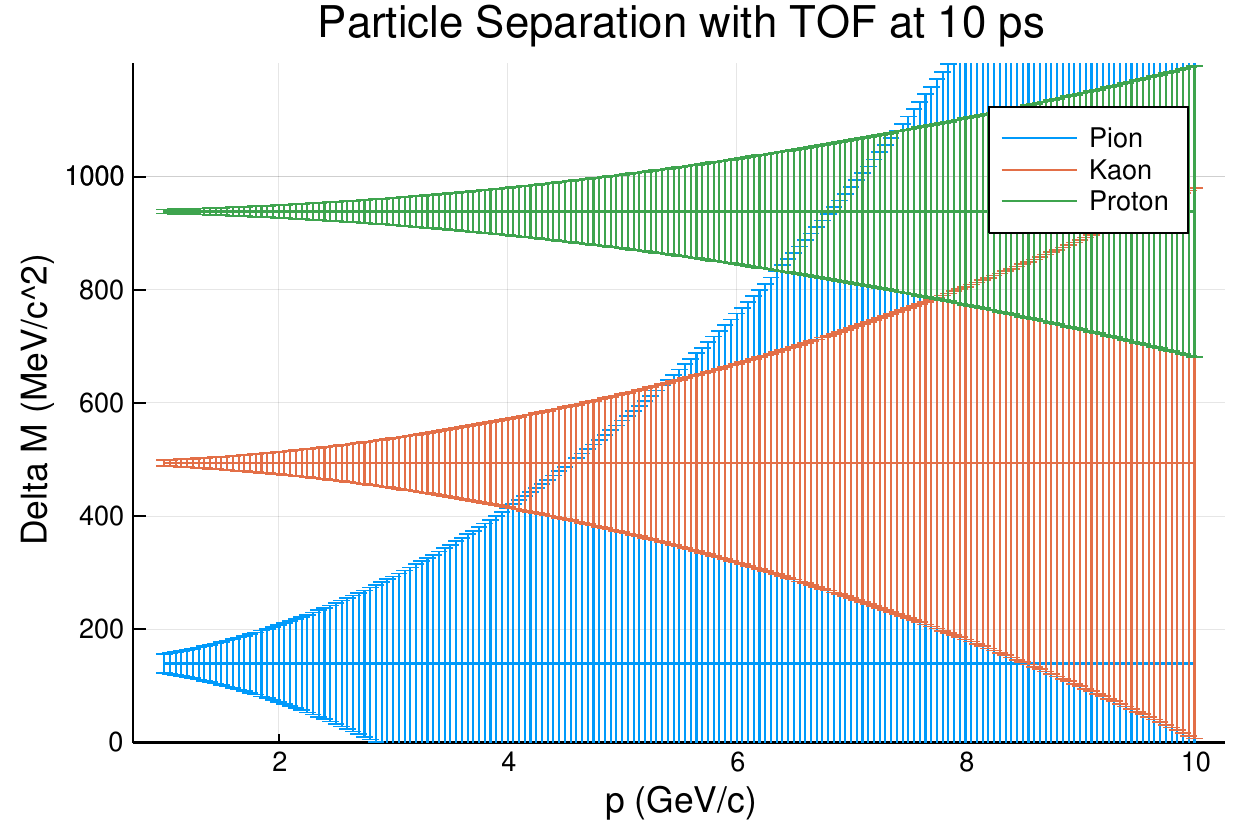}
\caption{Mass resolution for a TOF system with a performance of \SI{10}{\pico\second} in \SID \label{fig:Tofilc}}
 \end{center}
   \vspace{-0.7cm}
 \end{figure}

A timing layer as part of the tracking system or between tracker and ECAL could serve as a powerful Time-of-Flight system (TOF), where the physics reach needs 
to be further studied. It is already clear, that \SID would need to target a timing resolution in order of \SI{10}{\pico\second} which is roughly a factor three to five better than 
the resolution target by the timing layers of ATLAS~\cite{CERN-LHCC-2020-007} and CMS~\cite{CERN-LHCC-2017-027}.
A back-of-the-envelope calculation~\cite{TOFStudy} of a TOF layer located between the tracker and ECAL is shown in Figures~\ref{fig:Tofhllhc} and Figure~\ref{fig:Tofilc}.

\subsection{Particle ID}
Currently there is no dedicated Particle ID (PID) detector foreseen for \SID and a compelling physics case for a PID system besides timing-layer 
based solutions like a Time-of-Flight system would need to be demonstrated. Referring to Figures~\ref{fig:Tofhllhc} and Figure~\ref{fig:Tofilc} 
shows clearly that even with an excellent TOF system, a dedicated PID system is required, if there is a physics need for $\pi$/K separation 
with particle momenta greater than a few GeV/c.

\section{Conclusion}
\SID was originally conceived in the early 2000's and technology has since then made major steps forward. With the  realization of a linear collider becoming a real possibility, it is time to revisit the technology choices and update them accordingly. 
The replacement of the silicon strips and pad systems in the tracker and ECAL with MAPS is an obvious decision given recent advances. 

\clearpage

\bibliography{references.bib}

\end{document}